# Analysis of Wall Heat Flux of a Hypersonic Shock Wave / Boundary Layer Interaction with a Novel Decomposition Formula

Xiaodong Liu[1], Chen Li[1], Pengxin Liu[1], Qilong Guo[1], Xianxu Yuan[1*], Dong Sun[1†]

1 State Key Laboratory of Aerodynamics. Mianyang Sichuan 621000, China

**Abstract:** The generation mechanism of wall heat flux is one of the fundamental problems in supersonic/hypersonic turbulent boundary layers. A novel heat decomposition formula under the curvilinear coordinate was proposed in this paper. The new formula has wider application scope and can be applied in the configurations with grid deformed. The wall heat flux of an interaction between shock wave and the turbulent boundary layer over a compression corner is analyzed by the new formula. The results indicated good performance of the formula in the complex interaction region. The contributions of different energy transport processes were obtained. The contributions by the turbulent fluctuations e.g., Reynolds stresses and turbulent transport of heat flux, were significantly increased, while the processes by the mean profile e.g., molecular stresses and heat conduction, can be neglected. In addition, the pressure work is another contributor to the wall heat flux and the streamwise component works mainly in the shear layer and the reattachment point, while pressure in the wall-normal direction is concentrated in the vicinity of the reattachment point.

## Introduction

Shock wave/boundary layer interaction (SWBLI) is a common phenomenon in supersonic and hypersonic flows, and it is one of the most important issues that should be considered in the design of high-speed aircraft. It is well known that the peak heat flux could be extremely high caused by the SWBLI especially in hypersonic situations. Dolling[1] pointed out that the heat flux in the reattachment boundary layer can be 10-100 times of that in the upstream boundary layer of the flat plate, and even several times the stagnation point value, which seriously affects the flight safety of hypersonic vehicles. However, due to the lack of in-depth research on the heat flux in SWBLI, the prediction error of the current turbulence models is very large, which even reaches 100% of the experimental value [1].

---

* Corresponding author: yuanxianxu@cardc.cn
† Corresponding author: sundong0523@cardc.cn

To understand the generation mechanism of the wall heat flux in SWBLI, Lighthill [2] proposed a two-layer model in 1953, which holds the point that viscosity only works in the thin layer near the wall in the boundary layer. Neiland [3], Stewartson [4] and Mesiter [5] further put forward various three-layer structure models. Based on this, Brown [6] applied the three-layer structure models to the hypersonic shock wave interaction. Rizzetta [7] and Daniels [8] successfully predicted the heat flux in SWBLI by using the three-layer structure theory on the reattachment point. However, three-layer structure models cannot predict the wall heat flux in flows at low Reynolds numbers and with large separations well due to the assumptions in its derivations. After that, Fang and Bao [9] introduced an stagnation point-three-layer structure model theory when analyzing heat transfer characteristics in the reattachment region, which is used to investigate the influence of the size of the separation zone and the thickness of the boundary layer on the peak heat flux. Li and Bao [10] further proposed a compressible oblique stagnation point flow model to approximate the local flow and obtained a semi analytical and semi numerical estimation method of the distance between the reattachment point and the peak heat flux point. These theoretical methods help us to learn the generation mechanism of the heat flux, however, they cannot apply in complex situations due to the simplifications.

With the development of computer technology, direct numerical simulation (DNS) has gradually become an important means to study the complex turbulent flows. The first DNS computation in SWBLI problems is carried out by Adams et al. [11]. They simulated a supersonic flow at $Ma$=3.0, $Re_\theta$=1685 over a compression corner with the deflection angle of 18°. Wu and Martin [12] directly simulated the compression corners at $Ma$=2.9 and reproduced the low-frequency shock wave motion observed in the experiments. Volpiani and Bernardini et al. [13] studied the influence of wall temperature and found that wall heat flux would increase the length of the interaction region and the size of the separation bubble while wall cooling had an opposite effect. Zhu and Li et al. [14] studied the wall temperature effect through DNS, and found that, the separation bubble became larger with the increase of wall temperature. A semi-theoretical model that the separation bubble size is proportional to the 0.85th power of wall temperature is established. Compared to research on supersonic SWBLIs, DNS investigations under hypersonic conditions are rarely found. The difficulty in such occasions mainly comes from the complex coupling effect of strong shock wave and strong turbulence under hypersonic conditions, which makes the commonly used high-resolution scheme extremely unstable. Tong et al. [15] conduct the DNS on the shock interaction with a turbulence boundary layer at $Ma$=6. The strong intrinsic compressible effect is observed in the interaction region, and its impact on turbulent kinetic energy transport (TKE) is mainly reflected in the pressure-expansion term, while the expansion-dissipation term is less affected. Sun et al. [16] carried out DNS research on the shock wave/turbulent boundary layer in a hollow cylinder-flare configuration at $Ma$=6, and discussed the spanwise three-dimensional characteristics of the separation bubble structure in detail.

The abovementioned work has performed detailed research on the flow characteristics of the SWBLI, however, the generation mechanism of wall heat flux is barely seen. Following the idea of Renard et al [17], Sun et al. [18] made some integral transformations to the conservative equation of total energy, and quantitatively decomposed the wall heat flux into terms including the heat conduction, turbulent transport of heat, molecular stresses, the Reynolds stresses, etc. This heat flux decomposition method has been successfully used in the compressible boundary layer [18, 19]. To the author's knowledge, no heat flux decomposition research is reported in hypersonic SWBLI problems. On the one hand, the accuracy of the decomposition method can be reduced by the strong compressibility of shock wave. On the other hand, the more complicated configurations will be considered. In this paper, a novel decomposition formula of the wall heat flux was proposed. The new formula extended the flux decomposition from Cartesian coordinate system to curvilinear coordinate system. Importantly, new decomposition formula was applied to the heat flux analysis of hypersonic SWBLI, which quantitatively evaluated the influence of energy transport process on the wall heat flux during the interaction.

The work of this paper is organized as follows. In the first section, we describe numerical methods and DNS setups in brief. And the detailed derivation process of the heat flux decomposition formula in the curvilinear coordinate system is presented. The second section gives the analysis of the results, including the characteristics of the wall heat flux, the decomposition of wall heat flux at multiple positions, etc. Finally, conclusions are drawn in the last section.

# 1. Numerical methods and case setup

## 1.1 Governing equations

The governing equation adopted in this paper is the compressible Navier-Stokes equation, which is expressed as:

$$\frac{\partial \rho}{\partial t} + \frac{\partial \rho u_j}{\partial x_j} = 0$$

$$\frac{\partial \rho u_i}{\partial t} + \frac{\partial \rho u_i u_j}{\partial x_j} + \frac{\partial p \delta_{ij}}{\partial x_j} = \frac{\partial \sigma_{ij}}{\partial x_j} \qquad (1)$$

$$\frac{\partial \rho e}{\partial t} + \frac{\partial u_j(\rho e + p)}{\partial x_j} = \frac{\partial u_i \sigma_{ij}}{\partial x_j} + \frac{\partial q_j}{\partial x_j}$$

Where $u_1$, $u_2$ and $u_3$ denotes the streamwise, normal and spanwise velocity *u, v, w*, respectively. *p* represents pressure and $\rho$ is density. The expressions of total energy $\rho e$, molecular stress $\sigma_{ij}$, and heat flux $q_j$ are defined as:

$$\rho e = \frac{p}{\gamma - 1} + \frac{1}{2}\rho(u^2 + v^2 + w^2)$$

$$\sigma_{ij} = \mu\left[\left(\frac{\partial u_i}{\partial x_j} + \frac{\partial u_j}{\partial x_i}\right) - \frac{2}{3}\delta_{ij}\frac{\partial u_k}{\partial x_k}\right] \qquad (2)$$

$$q_j = \kappa \frac{\partial T}{\partial x_j}$$

The usual indicial notation is used. Prandtl number *Pr* is set as 0.72 and the specific heat ratio $\gamma$ is 1.4. The dynamic viscosity $\mu$ is obtained by using Sutherland's law, and $\kappa$ is the thermal conductivity coefficient. $Ma_\infty$ is the freestream Mach number and $Re_\infty$ is the Reynolds number, which is calculated according to the reference length $L_\infty$ of 1mm and the freestream parameter.

### 1.2 New Heat flux decomposition formula in curvilinear coordinate system

The wall heat flux of turbulent boundary layer is closely related to the energy transport process in the boundary layer. According to the idea of Renard [17], we propose a formula to decompose the heat flux coefficient by integrating the conservative equation of the total energy, which quantitatively decomposes the wall heat flow into the sum of the contributions of different energy transport terms such as the heat conduction, turbulent heat flux, molecular diffusion, molecular stresses, and the Reynolds stresses [18]. However, the heat flux decomposition formula is derived based on the Cartesian coordinate system, which is only suitable for simple configurations such as flat plates and channels and cannot applied to more complex configurations. In this paper, we further proposed a new heat flux decomposition formula in the general curvilinear coordinate system. The specific derivation process is given below.

The total energy conservation equation in Eq. (1) is averaged in time and spanwise to obtain:

$$\frac{\partial}{\partial t}\left[\overline{\rho}\left(\tilde{e}+\frac{\tilde{u}_i\tilde{u}_i}{2}\right)+\frac{\overline{\rho u_i''u_i''}}{2}\right]+\frac{\partial}{\partial x_j}\left[\overline{\rho}\tilde{u}_j\left(\tilde{e}+\frac{\tilde{u}_i\tilde{u}_i}{2}\right)+\frac{\overline{\rho u_j''u_i''}}{2}+\overline{p}\tilde{u}_j\right]$$
$$=\frac{\partial}{\partial x_j}\left[-q_{L,j}-\overline{\rho u_j''h''}+\overline{\sigma_{ji}u_i''}-\frac{\overline{\rho u_j''u_i''u_i''}}{2}\right]+\frac{\partial}{\partial x_j}\left[\tilde{u}_i\left(\overline{\sigma_{ij}}-\overline{\rho u_i''u_j''}\right)\right]$$
(3)

"-" denotes Reynolds average, and "~" denotes Favre average. Further, Eq. (2) can be abbreviated as:

$$\frac{\partial\overline{\rho}\tilde{E}}{\partial t}+\frac{\partial\overline{\rho}\tilde{u}_j\tilde{E}}{\partial x_j}+\frac{\partial\overline{p}\tilde{u}_j}{\partial x_j}=\frac{\partial Q_j}{\partial x_j}+\frac{\partial D_j}{\partial x_j}+\frac{\partial T_j}{\partial x_j}+\frac{\partial W_j}{\partial x_j} \qquad (4)$$

where $\tilde{E}$ is the average total energy; $Q_j$ is the heat flux composed of heat conduction and turbulent heat flux; $D_j$ is the molecular diffusion term; $T_j$ is the turbulent transport of TKE; $W_j$ is the work by molecular stresses and Reynolds stresses. Their specific expressions are:

$$\tilde{E}=\tilde{e}+\frac{\tilde{u}_i\tilde{u}_i}{2}+\frac{\widetilde{u_i''u_i''}}{2} \qquad (5)$$

$$Q_j=-q_{L,j}-\overline{\rho u_j''h''} \qquad (6)$$

$$D_j=\overline{\sigma_{ji}u_i''} \qquad (7)$$

$$T_j=-\frac{\overline{\rho u_j''u_i''u_i''}}{2} \qquad (8)$$

$$W_j=\tilde{u}_i\left(\overline{\sigma_{ij}}-\overline{\rho u_i''u_j''}\right) \qquad (9)$$

Transform the coordinate of Eq. (4) to a curvilinear coordinate system (ξ, η),

$$\begin{aligned}&J^{-1}\overline{\rho}\frac{D\tilde{E}}{Dt}\\&=\frac{\partial J^{-1}Q_\eta}{\partial\eta}+\frac{\partial J^{-1}D_\eta}{\partial\eta}+\frac{\partial J^{-1}T_\eta}{\partial\eta}\\&+\frac{\partial J^{-1}W_\eta}{\partial\eta}-\frac{\partial J^{-1}P_\eta}{\partial\eta}+\frac{\partial J^{-1}Q_\xi}{\partial\xi}\\&+\frac{\partial J^{-1}D_\xi}{\partial\xi}+\frac{\partial J^{-1}T_\xi}{\partial\xi}+\frac{\partial J^{-1}W_\xi}{\partial\xi}-\frac{\partial J^{-1}P_\xi}{\partial\xi}\end{aligned} \qquad (10)$$

Where:

$$\begin{aligned}Q_\xi&=Q_x\xi_x+Q_y\xi_y & Q_\eta&=Q_x\eta_x+Q_y\eta_y\\D_\xi&=D_x\xi_x+D_y\xi_y & D_\eta&=D_x\eta_x+D_y\eta_y\\T_\xi&=T_x\xi_x+T_y\xi_y & T_\eta&=T_x\eta_x+T_y\eta_y\\W_\xi&=W_x\xi_x+W_y\xi_y & W_\eta&=W_x\eta_x+W_y\eta_y\\PW_\xi&=\overline{p}\tilde{u}\xi_x+\overline{p}\tilde{v}\xi_y & PW_\eta&=\overline{p}\tilde{u}\eta_x+\overline{p}\tilde{v}\eta_y\end{aligned} \qquad (11)$$

Transform Eq. (10) to the relative coordinate system,

$$\begin{aligned}\xi_a&=\xi-U_{\xi,ref}t\\U_{\xi,a}&=U_\xi-U_{\xi,ref}\end{aligned} \qquad (12)$$

$U_{\xi,ref}$ is defined as the inverse velocity at the reference position $\eta_{ref}$. Move $Q_\eta$ to the left, everything else to the right, and multiply both sides by $U_{\xi,a}$

$$U_{\xi,a}\frac{\partial J^{-1}Q_{\eta,a}}{\partial \eta}=U_{\xi,a}J^{-1}\bar{\rho}\frac{D\tilde{E}_a}{Dt}-U_{\xi,a}\frac{\partial}{\partial \eta}[J^{-1}(D_{\eta,a}+T_{\eta,a}+W_{\eta,a}-P_{\eta,a})]$$
$$-U_{\xi,a}\frac{\partial}{\partial \xi}[J^{-1}(Q_{\xi,a}+D_{\xi,a}+T_{\xi,a}+W_{\xi,a}-P_{\xi,a})]\qquad(13)$$

Integrate Eq. (13) in the normal direction,

$$\int_0^{\eta_{ref}} U_{\xi,a}\frac{\partial J^{-1}Q_{\eta,a}}{\partial \eta}d\eta=\int_0^{\eta_{ref}}\frac{\partial J^{-1}Q_{\eta,a}U_{\xi,a}}{\partial \eta}d\eta-\int_0^{\eta_{ref}} J^{-1}Q_{\eta,a}\frac{\partial U_{\xi,a}}{\partial \eta}d\eta$$
$$=(J^{-1}Q_{\eta,a}U_{\xi,a})_{\eta_{ref}}-(J^{-1}Q_{\eta,a}U_{\xi,a})_0-\int_0^{\eta_{ref}} J^{-1}Q_{\eta,a}\frac{\partial U_{\xi,a}}{\partial \eta}d\eta \qquad(14)$$
$$=J^{-1}Q_{\eta,a}U_{\xi,ref}\big|_{\eta=0}-\int_0^{\eta_{ref}} J^{-1}Q_{\eta,a}\frac{\partial U_{\xi,a}}{\partial \eta}d\eta$$

Further, restore to the absolute coordinate system

$$Q_w=\frac{1}{C_\eta}\int_0^{\eta_{ref}} J^{-1}Q_\eta\frac{\partial U_\xi}{\partial \eta}d\eta+\frac{1}{C_\eta}\int_0^{\eta_{ref}} (U_\xi-U_{\xi,ref})J^{-1}\bar{\rho}\frac{D\tilde{E}}{Dt}d\eta$$
$$+\frac{1}{C_\eta}\int_0^{\eta_{ref}} J^{-1}(D_\eta+T_\eta+W_\eta-P_\eta)\frac{\partial U_\xi}{\partial \eta}d\eta \qquad(15)$$
$$-\frac{1}{C_\eta}\int_0^{\eta_{ref}} (U_\xi-U_{\xi,ref})\frac{\partial}{\partial \xi}[J^{-1}(Q_\xi+D_\xi+T_\xi+W_\xi-P_\xi)]d\eta$$

Where $C_\eta=J^{-1}U_{\xi,ref}\sqrt{\eta_x^2+\eta_y^2}$.

So far, the heat flux decomposition formula in the curvilinear coordinate system si obtained. By substituting Eqs. (5)-(9) into Eq. (15), the contributions of the heat conduction $Q_L^{\xi/\eta}$, turbulent heat transport $Q_T^{\xi/\eta}$, molecular diffusion $Q_D^{\xi/\eta}$, TKE transport $Q_K^{\xi/\eta}$, molecular stress work $Q_{MW}^{\xi/\eta}$, Reynolds stress work $Q_{RW}^{\xi/\eta}$, pressure work $Q_{PW}^{\xi/\eta}$ and convection term in the directions of ξ and η can be evaluated respectively. In the following analysis, the heat flux decomposition formula (Eq. (15)) derived in curvilinear coordinate system is used to investigate the wall heat flux of SWBLI. On the one hand, the accuracy and reliability of the formula can be verified; on the other hand, the heat flux generation mechanism under SWBLI can be further understood.

## 1.3 Numerical methods and case setup

An in-house code is employed to perform the DNS study. This code has been applied in many DNS calculations of compressible turbulent cases[20-21], and the accuracy and robustness have been well validated. In this paper, a hybrid high-order scheme is adopted to discretize the inviscid fluxes. An improved Ducros detector [22] is used to divide the flow field into a smooth region and a discontinuous region. In the discontinuous region, the NND scheme with TVD property is adopted [23, 24], while in the smooth region, the optimized low dissipation WENO scheme is selected [25]. The viscous terms are discretized by the sixth-order central scheme, and the third-order TVD Runge-Kutta method [26] is adopted as the temporal algorithm.

The incoming flow Mach number is 6, the unit Reynolds number is $1.2 \times 10^7$, the incoming flow temperature is 65 K, and the wall temperature is 305.5 K. The sketch of the computational domain and boundary conditions are presented in Fig. 1. The angle of the compression corner is 33°. The length $L_{x1}$ of the plate in front of the corner is 257 mm, the length $L_{x2}$ of the ramp behind the corner is 116 mm, the normal height $L_y$ of the inlet position is 35 mm, and the spanwise width $L_z$ is 28 mm. The computational domain is discretized with $N_x \times N_y \times N_z = 1151 \times 320 \times 149$ points. Grid points are equally spaced in the spanwise direction, and the grid spacing is $d_z^+=3.1$. In the streamwise direction, the resolution at the inlet position is $d_x^+=7.2$., and the points are properly refined at the corner. The grid points in the normal direction are clustered near the wall and the first layer height is $d_y+=0.32$. The boundary conditions are set as: the inlet uses a digital filtering method to provide synthetic turbulence [27], an isothermal nonslip condition is applied at the wall, the periodical conditions are set on both sides in the spanwise direction, and the outlet and upper boundary are set as non-reflection conditions [28].

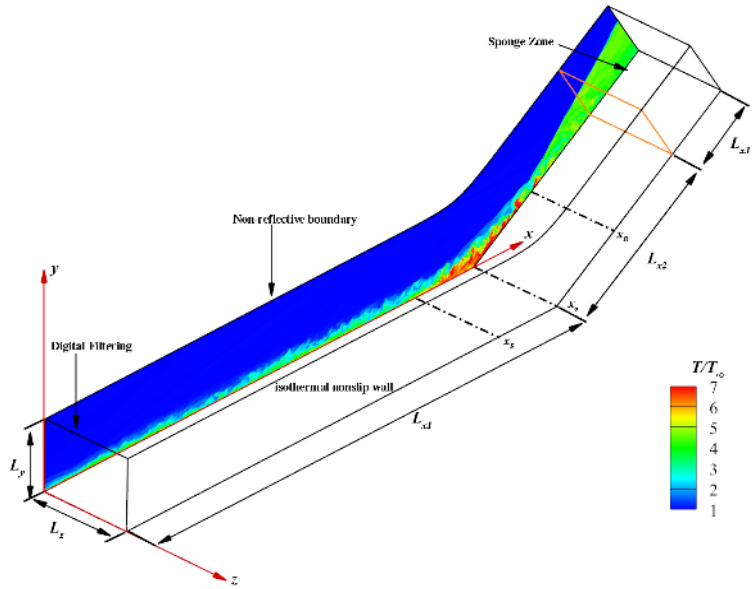

Fig. 1 Computaional domain and boundary condition settings

## 2. Result analysis

### 2.1 DNS data analysis

Fig. 2 shows the distribution of the van Driest transformed streamwise velocity, where the classic law in the viscous sublayer and logarithmic region is observed. The results agree well with that reported in DNS of hypersonic boundary layers in the literature [29,30]. The normalized Reynolds stresses are presented in Fig. 3. Compared to the results in literature [31-33], good agreement in the outer region at $y/\delta > 0.1$ is achieved while some differences are observed in the inner layer at $y/\delta < 0.1$. A possible reason for the differences should be that Morkorvin hypothesis is invalid in hypersonic flows. Generally, the turbulence provide by the digital filtering method at the inlet in the present simulation is acceptable.

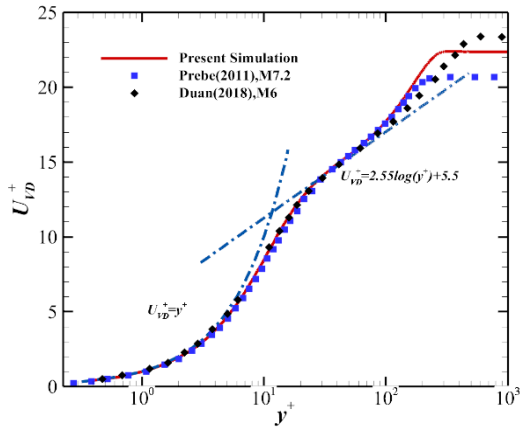

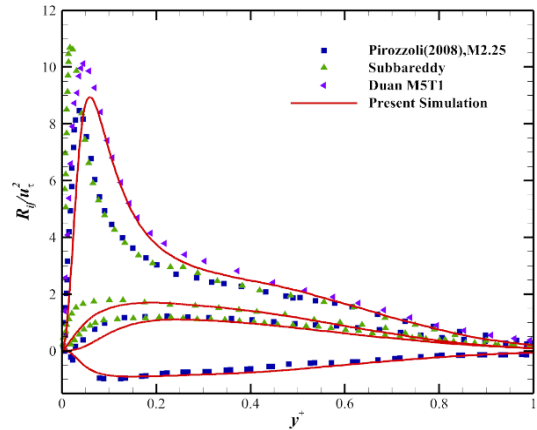

Fig. 2 The van Driest transformed mean streamwise velocity

Fig. 3 The Reynolds stresses normalized by the friction velocity

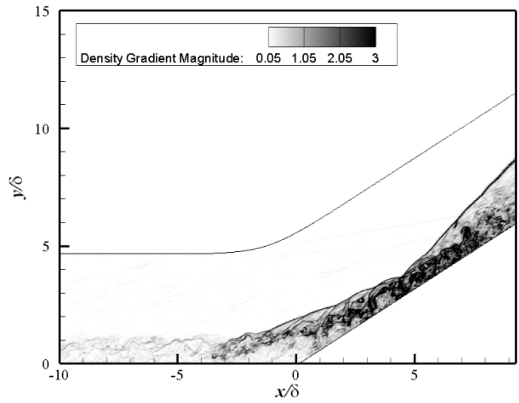

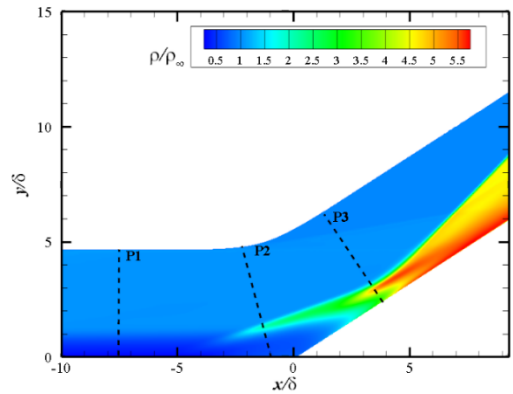

Fig. 4 the instantaneous contours of the magnitude of the density gradients

Fig. 5 The contours of the mean density in the interaction

To give an overall impression of the flow characteristics of SWBLI obtained by DNS, instantaneous density gradient and mean density are depicted in Fig. 4 and Fig. 5, respectively. As presented in Fig. 4, the intensity of density gradient is enhanced after the separation shock wave. Meanwhile, compression waves are formed above the separation bubble. After entering the downstream interaction region, the shock wave intensity is further enhanced, reflecting the strong interaction with the boundary layer. In addition, the flow structures near the outlet experience a rapid decay after the shock wave, which is caused by the buffer zone set in the outlet. Fig. 5 shows the contour of time and spanwise averaged density. Typical structures such as the separation shock, the reattachment shock, and the separation bubble are clearly presented. The density does not increase significantly in the separation bubble area but it mainly concentrated in the area near the separation shock wave and after the reattachment shock wave. In Fig. 5, three typical locations are selected for the wall heat flux analysis, which are marked as P1, P2 and P3, respectively, where P1 is located in the upstream boundary layer, P2 is located in the separation bubble, and P3 is located after the reattachment.

Fig. 6 presents the streamwise distribution of mean wall pressure, skin friction and heat flux. The pressure is normalized by the freestream pressure, the skin friction is defined as: $C_f = \frac{2\tau_w}{\rho_\infty U_\infty^2}$, and the heat flux is defined as: $C_h = \frac{q_w}{\rho_\infty U_\infty} C_p (T_r - T_w)$. Here, "S" and "R" are marked to represent the separation point and the reattachment point. An obvious pressure plateau is found in the separation zone. Pressure and skin friction are increased rapidly after the reattachment, follows by a slow decrease after attaining the peak value. Similarly, there is a plateau in the heat flux distribution in the separation zone. And a heat flux peak appears after the rapid rise, in which the peak position is located between the reattachment point and the skin friction peak. To further verify the reliability of the present results, the Reynolds analogy factor RAF = $2C_h/C_f$, the relationship between skin friction and heat flux, is introduced. In the upstream turbulent boundary layer, RAF is a constant and generally close to one. In the present study, RAF≈1.2 is obtained, which agrees well with that reported by Priebe and Martin in the DNS investigations of a hypersonic turbulent boundary layer [29]. After reattachment, RAF decreases sharply and gradually approaching 1, which is related to the relaxation recovery process of the boundary layer after the interaction. The classic analogy of heat flux and pressure is also verified. Neumann et al. [35] proposed the relationship as $C_{h,pk}/C_{h,fp}=(P_{pk}/P_{fp})^n$, where the subscript "pk" is the peak value, and the subscript "fp" represents the upstream flat plate, $n$=0.5 in laminar flow and $n$=0.8 in turbulent flow. In the present study, $n$ is 0.78, which is in a good agreement with the empirical result.

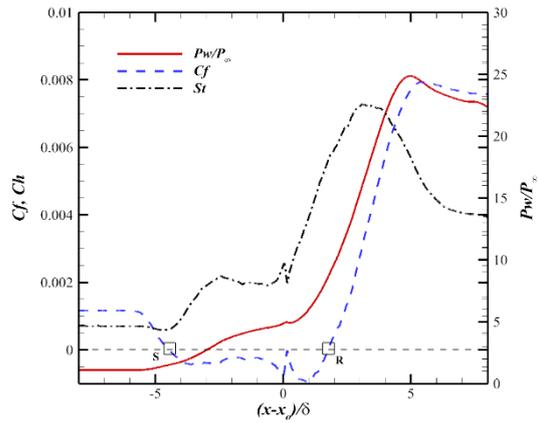 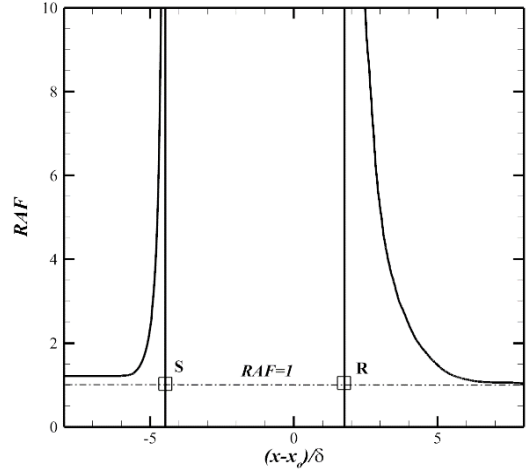

Fig. 6 Spanwise and time averaged pressure, friction coefficients and heat flux

Fig. 7 Reynolds analgoy factor along the streamwise direction

## 2.2 Statistical characteristics of wall heat flux

In addition to the analysis of mean variables, this section will analyze the statistical characteristics of the wall heat flux, including the probability density function (PDF), power spectral density (PSD) and two-point correlation function.

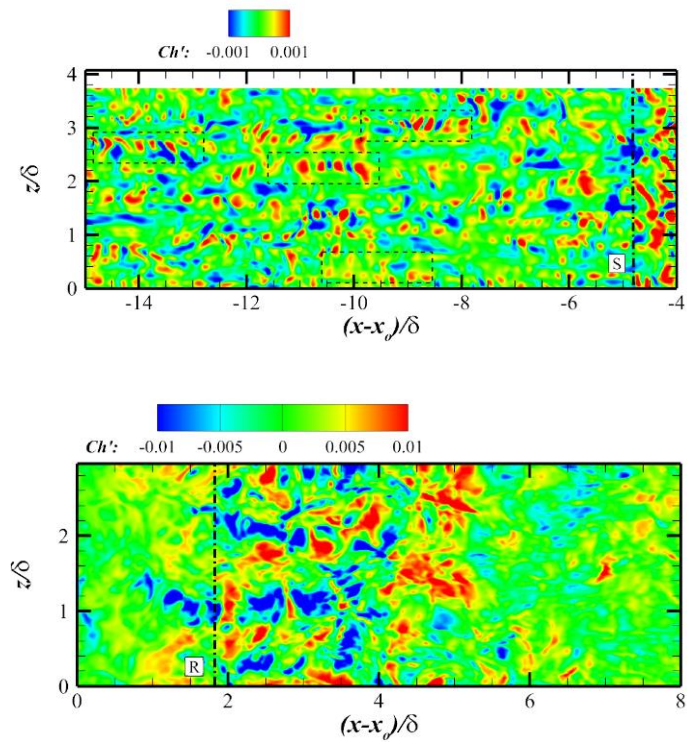

Fig. 8 Instantaneous contours of the fluctuations heat flux coefficient before (upper) and after (lower) the interaction

Fig. 8 reports contours of the instantaneous heat flux fluctuations before and after the interaction. Unlike that occurs in the skin friction, no streak structures are found from Fig.8(a) in the heat flux fluctuation. Instead, irregular heat flux spots are frequently observed. The heat flux on the flat plate witnesses positive and negative alternating spots before the interaction, which are marked with a black dashed box. And Yu[36] explained this traveling wave structure with the dilatation structures by Helmholtz decomposition. After the interaction, the traveling wave structure is destroyed. Then the heat flux began to fluctuate violently within the thickness range of three boundary layers after the reattachment point, after which the fluctuation intensity is decreased rapidly.

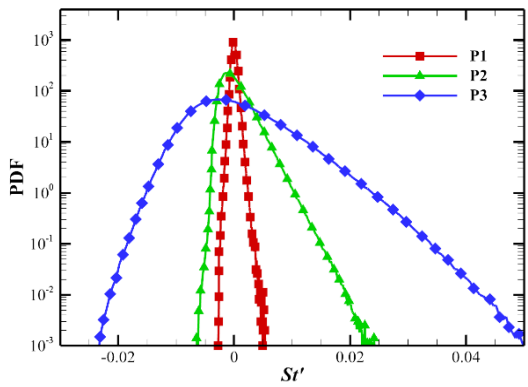 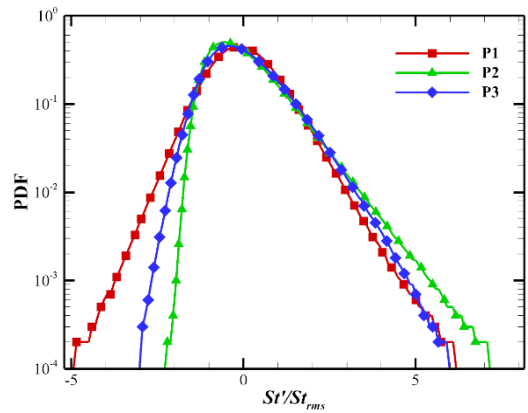

Fig. 9 Probility densty function of the fluctuations of the heat flux coefficients

Fig. 10 Probility densty function of the fluctuations of the heat flux coefficients normalized by the r.ms. of the coeffcients

To assess the distribution of heat flow fluctuation intensity, the PDF of heat flux at P1-P3 before and after the interaction is calculated. As shown in FIG. 9, asymmetric PDF is found at the three selected stations, and the asymmetry becomes more obvious when moving downward. In addition, the maximum and minimum values of heat flux fluctuation gradually increase along the streamwise direction. Specifically, the minimum value of heat flux fluctuation at P3 is 6 times that at P1, and the maximum value increases to more than 10 times, indicating that the growth of heat flux fluctuation is extremely fast under hypersonic conditions. In Fig. 9, the root mean square of the heat flux density is used to normalize the heat flux fluctuation. In the positive branch, the distributions of PDF collapse together and the negative branch, apparent differences can be observed.

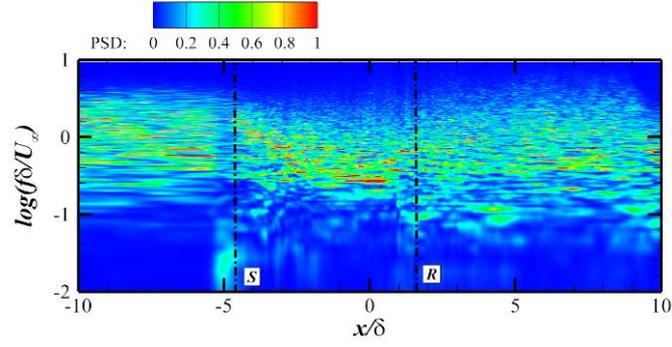

Fig. 11 Power spectral density of the fluctuations of the wall pressure

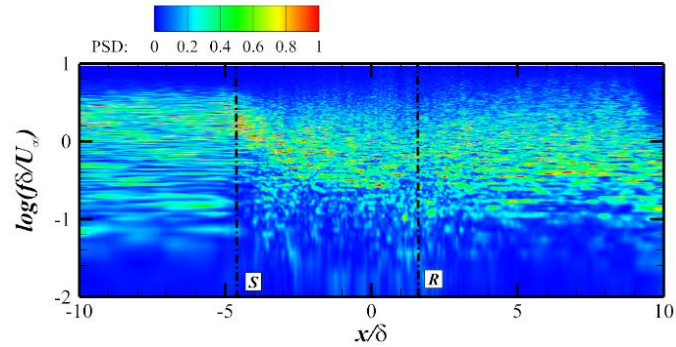

Fig. 12 Power spectral density of the fluctuations of the wall heat flux

The spectral characteristics of heat flux are further discussed. In Fig. 11, the PSD of the pressure fluctuation is presented. The low-frequency frequency appears at the separation point, which is $f\delta/U_\infty \approx 0.02$, consistent with the low-frequency frequency range given in the literature [38-39]. Fig. 12 further shows the PSD of heat flux fluctuation. The low-frequency signal at the separation point disappears, indicating that heat flux fluctuation is characterized by the medium-high frequency rather than low-frequency.

The scale characteristics of the heat flux fluctuation are further calculated. In Fig. 13, the two-point correlation of heat flux fluctuation before and after interaction is given, and the formula is

$$C_{St'St'}(dx,dz) = \frac{\overline{St'(x,y,z)St'(x+dx,y,z+dz)}}{\sqrt{\overline{St'(x,y,z)^2}\,\overline{St'(x+dx,y,z+dz)^2}}}$$

The contour is the correlation coefficient of heat flux fluctuation on the flat plate before the interaction, and the solid black line is the results after the interaction. The structure is changed significantly. Before the interaction, the correlation function presents a streak structure, with a circular spotted main structure presented. After the interaction, the overall shape of the coherence coefficient tends to be round, where the spanwise scale increases significantly. The transformation of correlation function is consistent with the shape change process of instantaneous heat flux fluctuation in Fig. 8.

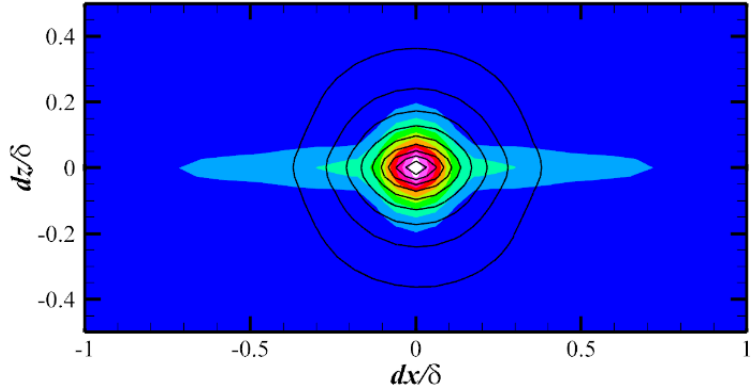

Fig. 13 two-point correlation function of the heat flux before (contours) and after (lines) the interaction

## 2.3 Decomposition of wall heat flux

In this section, the new heat flux decomposition formula in the curvilinear coordinate system is used to decompose the heat flux in a hypersonic compression ramp. The decomposition results of the heat flux $Q_{w,decomp}$ following Eq. (15) is given in Table 1, together with the relative error compare to the local mean heat flux. The decomposition results are accurate, for example, the error at P1 in the upstream turbulent boundary layer is only 0.05%. The decomposition error at P2 in the separation bubble is 4.84%. As P2 is located near the corner, where the grid is deformed, applying the original decomposition [18] based on the Catesian coordinate system is difficult. At P3, the error is 4.95%. Considering the complexity of the flow field in the interaction region, the errors of P1-P3 is acceptable. Therefore, the decomposition method of the curvilinear coordinate system derived in this paper still has a sufficient accuracy when dealing with the complex SWBLI problems.

Table 1 Decomposition of the heat flux at different streamwise positions

|    | $Q_{w,decomp}$ | $Q_{w,avg}$ | Error(%) |
|----|----------------|-------------|----------|
| P1 | 1.395E-4       | 1.395E-4    | 0.05     |
| P2 | 4.099E-4       | 3.900E-4    | 4.84     |
| P3 | 1.354E-3       | 1.421E-3    | 4.95     |

Fig. 14 shows the contributions of different energy transport processes to the wall heat flux at P1, $rQ_L^{\xi/\eta}$ represents the ratio of the heat conduction contribution to the average heat flux, $rQ_T^{\xi/\eta}$ represents the ratio of the turbulent heat transport contribution to the average heat flux, similarly, $rQ_D^{\xi/\eta}$, $rQ_K^{\xi/\eta}$, $rQ_{MW}^{\xi/\eta}$, $rQ_{RW}^{\xi/\eta}$, $rQ_{PW}^{\xi/\eta}$ and $rQ_C$ represent the ratio of molecular diffusion, turbulent kinetic energy transport, molecular stresses work, Reynolds stresses work, pressure work term and convective transport to the wall heat flux contribution to the average heat flux, respectively. And the sum of contribution in the $\xi$ direction is denoted as $\sum rQ_\xi$. It can be seen from Fig. 14 that molecular stresses work $rQ_{MW}^\eta$ and Reynolds stresses work $rQ_{RW}^\eta$ are two major contributors in the upstream flat plate, and the contribution of the Reynolds stresses work is larger than that of the molecular stresses work.

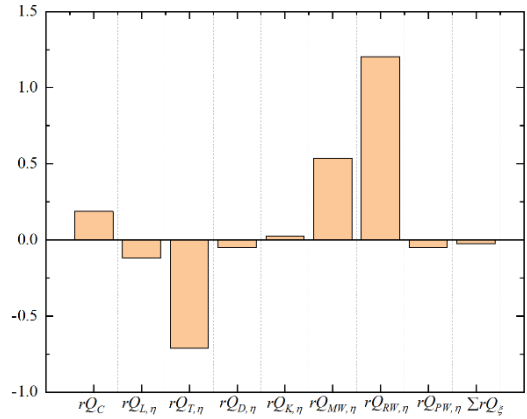

Fig. 14 Contributions of different energy transport process to the wall heat flux at P1

Turbulent heat flux transport term $rQ_T^\eta$ is the main negative contribution of heat flux, and the heat flux generated near the wall is transferred to the outer boundary layer through turbulent pulsation. In addition, the heat conduction term $rQ_L^\eta$ is also a negative heat flux contribution term, which is very small compared with turbulent heat flux transport. And the convection transport term $rQ_C$ is also a significant contribution, mainly due to the nonuniform distribution of total energy in the streamwise direction. Finally, the contribution of molecular diffusion term $rQ_D^{\xi/\eta}$ and turbulent kinetic energy transport term $rQ_K^{\xi/\eta}$ is very small, and the contribution of energy transport process $\sum rQ_\xi$ in the streamwise direction is also negligible.

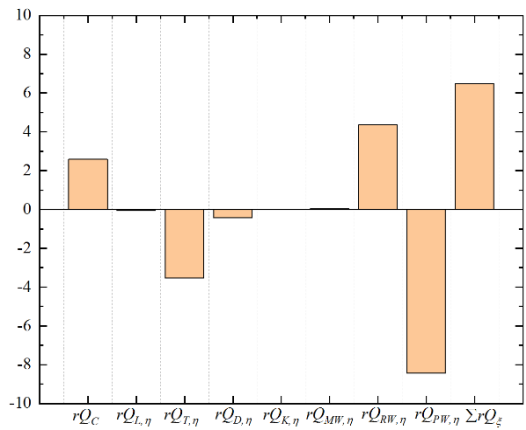

Fig. 15 Contributions of different energy transport process to the wall heat flux at P2

Figure 15 shows the decomposition at P2. The first thing to note is that the ratios of different contributions to the average heat flux vary greatly. The contribution of Reynolds stresses work to the wall heat flux is about five times of the average heat flux, and the contribution of turbulent heat flux transport is also increased from 0.8 times of the original average heat flux to three times. The reason for these two increases is that the amplitude of velocity fluctuation and temperature fluctuation increase significantly after the shock wave. Secondly, molecular stress work and heat conduction do not increase significantly, far less than the contributions of turbulent heat flux transport and Reynolds stresses work, which means that the average amount will increase behind the shock wave, but the increase amplitude is much smaller than the fluctuation amount increase. Thirdly, we notice that the pressure work term has a large negative contribution to the wall heat flux, with the amplitude reaching nine times of the average heat flux, and the energy transport in the streamwise direction also has a large positive contribution. The above situation is a new feature to the flat plate turbulence, which is caused by the influence of SWBLI and the nonuniform streamwise direction caused by the compression corner, which will be further discussed later.

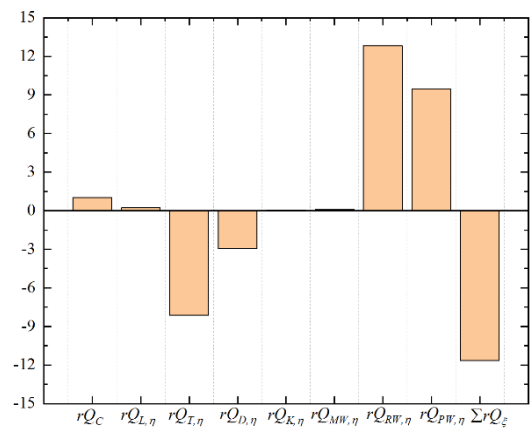

Fig. 16 Contributions of different energy transport process to the wall heat flux at P3

Fig. 16 shows the contribution of different energy transport to heat flux near the reattachment point of the SWBLI at P3. Compared with P2, the amplitude of turbulent heat flux transport term and Reynolds stresses work term are further increased, while the contributions of heat conduction and molecular stresses work are still small and negligible. The contribution of the convection term is reduced, because P3 is already on the inclined flat plate, and the gradient change of the total energy is reduced. Although the contribution of normal pressure work and streamwise direction energy transport is still large, but the contribution of pressure work is positive and the contribution of streamwise energy transport is negative.

The different streamwise energy transport contributions at P2 and P3 are discussed below. At P2, the contributions of turbulent heat flux transport, turbulent kinetic energy transport, Reynolds stresses work and pressure work are the most significant, while the contributions of other terms are small and can be ignored. The results show that the most significant increase of streamwise energy transport is the turbulent fluctuation terms and the pressure work term. The SWBLI on turbulence fluctuations will significantly increase the wall heat flux in both normal direction and streamwise direction. At P3, the contribution to wall heat flux is small except for the streamwise direction Reynolds stresses work and the pressure work. This is because P3 is located on an inclined flat plate and the flow gradient decreases.

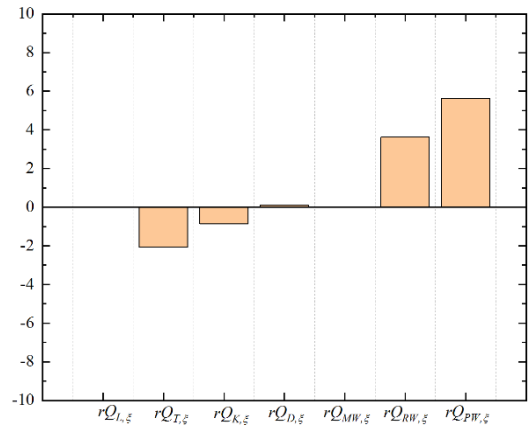

Fig. 17 Contributions of different energy transport process in the streamwise direction to the wall heat flux at P2

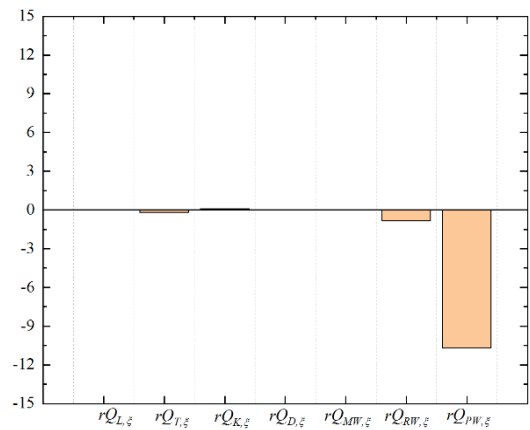

Fig. 18 Contributions of different energy transport process in the streamwise direction to the wall heat flux at P3

In order to analyze the contribution of the pressure work term, the contour of the streamwise pressure work $PW_\xi$ and the normal pressure work $PW_\eta$ are calculated. It can be seen from Fig. 19 that the streamwise pressure work is mainly concentrated near the shear layer above the separation bubble and behind the reattachment point, especially behind the reattachment point, the pressure work reaches the maximum value and can decrease rapidly after maintaining a long streamwise distance. Black dotted lines are used to mark positions P1-P3 in the figure, and the normal distribution curve of the streamwise pressure work is obtained, as shown in Fig. 20. The streamwise pressure work on the upstream flat plate concentrates on the position very close to the wall and then remains unchanged. Because the pressure on the flat plate is approximately unchanged along the normal direction, the streamwise pressure work mainly comes from the distribution of streamwise velocity. The streamwise pressure work at P2 increased for a long time and then declined rapidly. The pressure at P3 increased rapidly and declined after reaching the peak. It can be seen from Eq. (15) that the contribution of work done by streamwise pressure to the wall heat flux is:

$$Q_{PW,\xi} = \frac{1}{C_\eta} \int_0^{\eta_{ref}} (U_\xi - U_{\xi,ref}) \frac{\partial J^{-1} P_\xi}{\partial \xi} d\eta.$$

The contribution of the streamwise pressure work is mainly determined by the product of the term $U_\xi - U_{\xi,ref}$ and the derivative of $P_\xi$ along the streamwise direction. And the term $U_\xi - U_{\xi,ref}$ always is negative in the boundary layer, Therefore, the streamwise pressure work as shown in Fig. 16-18 contributes little to P1, positive to P2, and negative to P3.

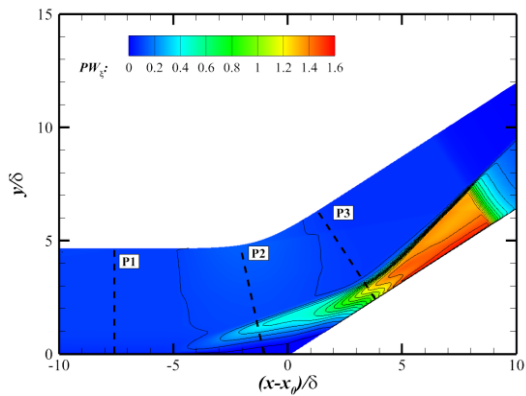 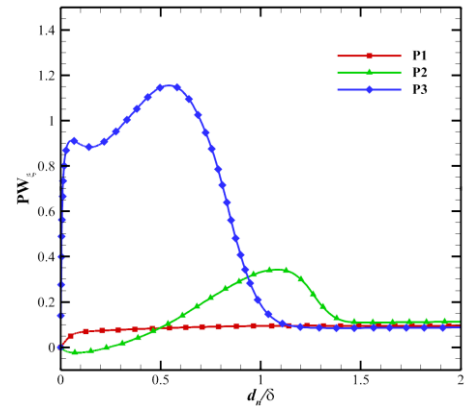

Fig. 19 contours of the work by pressure in the streamwise direction

Fig. 20 Normal distribution of the work by pressure in the streamwise direction at P1-P3.

Fig. 21 shows the distribution contour of the normal pressure work. It can be seen from the figure that the amplitude of work done by normal pressure is the largest near the reattachment region and the smaller near the separation bubble, which is different from the distribution of streamwise pressure work. Fig. 22 shows the distribution of normal pressure work at p1-P3 along the normal direction. It can be seen from the figure that the contribution of normal pressure work on the upstream flat plate is small and can be ignored, while the amplitude of normal pressure work at P2 position is positive and that at P3 position is negative. According to Formula (15), the contribution of normal pressure to heat flux can be expressed as:

$$Q_{PW,\eta} = -\frac{1}{C_\eta} \int_0^{\eta_{ref}} J^{-1} P_\eta \frac{\partial U_\xi}{\partial \eta} d\eta.$$

Thus, the contribution of normal pressure work at P2 to the wall heat flux is negative but that of normal pressure work at P3 is positive.

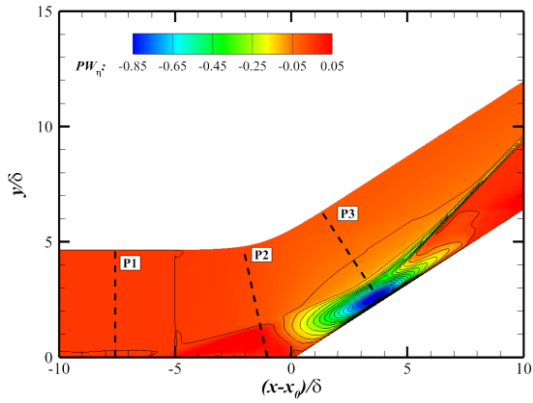 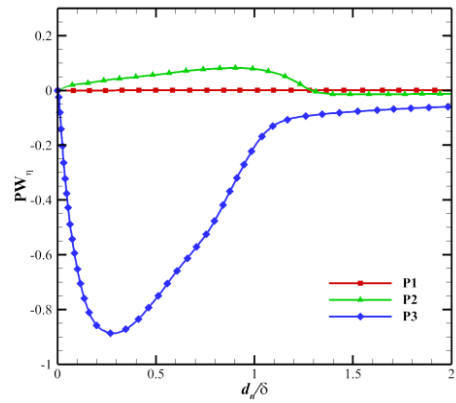

Fig. 21 contours of the work by pressure in the normal direction

Fig. 22 Normal distribution of the work by pressure in the normal direction at P1-P3.

## 3 Conclusion

A heat flux decomposition method in curvilinear coordinate system is developed to understand the wall heat flux features in hypersonic SWBLI flows. Statistical characteristics and energy transport contributions of wall heat flux in the compression corner at Mach number of 6 are studied by direct numerical simulation. Conclusions can be drawn as follows.

(1) In terms of statistical characteristics of the heat flux fluctuation, it is significantly increased by SWBLI. The fluctuating is featured as medium and high frequent, and no low-frequency components are observed. In addition, the statistical structures change from streaks to spots after the interaction, with a greatly increased spanwise scale.

(2) As for the heat flux decomposition, selected stations before and after the SWBLI are analyzed through the heat flux decomposition method. The accuracy and reliability of this technology are well validated first. Turbulent fluctuation dominated energy transportation, including the work by Reynolds stresses and turbulent transport of heat, is significantly increased after the interaction. By contrast, the contribution dominated by the average profile, such as the heat conduction and the molecular stresses work, is negligible.

(3) The work by pressure has significant contribution to the wall heat flux in the SWBLI flow. The streamwise component works mainly in the shear layer and after the reattachment point, while pressure in the wall-normal direction is concentrated in the vicinity of the reattachment point.

## Acknowledgement

This work was supported by the National Key Research and Development Program of China (Grant No. 2019YFA0405201), the National Natural Science Foundation of China (Grant No. 92052301, 11902345), and the National Numerical Windtunnel Project.